\documentstyle[twocolumn,aps,prl,epsf]{revtex}
\begin{document}

\twocolumn[\hsize\textwidth\columnwidth\hsize\csname
@twocolumnfalse\endcsname

\title{Elastic Properties of Single-Wall Nanotubes}

\author{E.~Hern\'{a}ndez$^1$\thanks{To whom correspondence should be addressed.
E-mail address: ehe@boltzmann.fam.cie.uva.es}, C.~Goze$^2$, P.~Bernier$^2$
and A.~Rubio$^1$} 
\address{$^1$~Departamento de F\'{\i}sica Te\'{o}rica, 
        Universidad de Valladolid, 
        Valladolid 47011, Spain \\
        $^2$~Groupe de Dynamique des Phases Condens\'{e}es, 
         Universit\'{e} Montpellier II, 34090 Montpellier, 
        France}

\maketitle

\maketitle

\begin{abstract}
We report results of theoretical studies on the elastic properties of
single-wall nanotubes of the following compositions: C, BN, $\mbox{BC}_3$,
$\mbox{BC}_2\mbox{N}$ and $\mbox{C}_3\mbox{N}_4$. These studies have been
carried out using a total energy, non-orthogonal tight-binding parametrisation
which is shown to provide results in good agreement both with calculations
using higher levels of theory and the available experimental data. 
Our results predict
that of all types of nanotubes considered, carbon nanotubes have the highest
Young's modulus. We have considered tubes of different diameters, ranging
from 0.5 to 2~nm, and find that in the limit of large diameters the 
mechanical properties of nanotubes approach those of the corresponding 
flat graphene-like sheets.
\end{abstract}

]

\narrowtext

The discovery of $\mbox{C}_{60}$ and fullerenes~\cite{kroto}
in the mid 80's was soon followed by the observation of 
nanotubes~\cite{nanotube_reviews}, first reported by Iijima~\cite{iijima}
in 1991. Since then nanotubes have been the focus of attention of a 
growing scientific community, attracted to them by their many 
interesting properties, such as their structure, electrical 
conductivity and mechanical properties, as well as by their large
potential for practical applications. Two types of nanotubes exist:
those originally observed by Iijima~\cite{iijima} were multi-wall
nanotubes (MWNT's), formed by concentric shells of apparently seamless
cylinders of graphene, having a separation between them similar to that
in graphite. More recently, single-wall nanotubes (SWNT's) have also been
synthesized. As their name indicates, these consist of a single seamless
cylinder of graphene~\cite{nanotube_reviews}.

Soon after the discovery of carbon nanotubes it was proposed that other
compounds forming graphite-like structures, such as BN~\cite{rubio:corkill:cohen}, 
$\mbox{BC}_3$~\cite{miyamoto}, $\mbox{BC}_2\mbox{N}$~\cite{BCN}, 
and CN~\cite{CN}, could also form nanotubular structures.
Indeed BN~\cite{bn_synthesis,loiseau,terrones}, 
$\mbox{BC}_3$ and $\mbox{BC}_2\mbox{N}$~\cite{bc3_synthesis} have now
been synthesized, though the actual structure of $\mbox{BC}_2\mbox{N}$ tubes
seems to correspond to concentric shells of C and BN in a 'sandwich'
structure~\cite{segregation}. Other tubular structures formed by heavier
element compounds have  been predicted, such as GaSe~\cite{cote}, and
synthesized, like $\mbox{WS}_2$ and $\mbox{MoS}_2$~\cite{tenne}.

In this paper we focus our attention on the structural, energetic and 
mechanical properties of single-wall carbon and composite nanotubes.
We perform a systematic study of these systems using Tight-Binding 
total energy methods~\cite{tb:review}, as well as 
first-principles~\cite{pw:review}
calculations. Some of the results reported here have
already appeared in published form~\cite{prl}, but we also provide 
previously unpublished results for the $\mbox{C}_3\mbox{N}_4$ 
nanotubes.

The structure of this paper is as follows. In Sec.~\ref{sec:experimental}
we review the available experimental data on the mechanical properties of 
nanotubes, while Sec.~\ref{sec:theoretical} is devoted to discussing previous
theoretical work. In Sec.~\ref{sec:model_and_calcs} we describe the 
models used in our work and the calculations carried out in order to 
address the issues discussed here. Then, in Sec.~\ref{sec:results} we 
discuss our results and conclusions.

\section{Review of Experimental Results}
\label{sec:experimental}

There is a growing body of experimental evidence 
indicating that carbon nanotubes (both
MWNT's and SWNT's) have extraordinary mechanical properties. There are many
direct observations of the large bending 
flexibility~\cite{iijima:bernholc,falvo,JPC} of nanotubes, which provide 
evidence of their capability to sustain large strains 
without evidence of collapse or failure. However, the technical difficulties
involved in the manipulation of these nano-scale structures makes the direct
determination of their mechanical properties a rather challenging task.
In spite of these difficulties, a number of experimental measurements of 
the Young's modulus of nanotubes have been reported. The first such study was
that of Treacy {\em et al.\/}~\cite{treacy}, who correlated the amplitude
of the thermal vibrations of the free ends of anchored nanotubes as a function
of temperature with the Young's modulus. Regarding a MWNT as a hollow cylinder
with a given wall thickness, one can obtain a relation between the amplitude
of the tip oscillations in the limit of small deflections, and the Young's
modulus. Having quantified the amplitude of those oscillations by means of
careful TEM observations of a number of nanotubes, Treacy {\em et al.\/} 
were able to obtain an average value of 1.8~TPa for the Young's modulus, though
there was significant scatter in the data (from 0.4 to 4.15~TPa for individual
tubes). Thus this number is subject to large error bars, but it is
nevertheless indicative of the exceptional axial stiffness of these materials.

More recently Krishnan {\em et al.\/}~\cite{krishnan} have reported studies
on SWNT's using the same technique. A larger sample of nanotubes was used, and 
a somewhat smaller average value was obtained, $Y= 1.25$~TPa, closer to the
expected value for graphite along the basal plane.

This technique has also been used by Chopra and Zettl~\cite{zettl} to estimate
$Y$ for BN nanotubes. Their results indicate that these composite tubes are
also exceptionally stiff, having a value of $Y$ around 1.22~TPa, very close
to the value obtained for carbon nanotubes.

Another way to probe the mechanical properties of nanotubes has been 
described by Wong {\em et al.\/}~\cite{lieber}, who have used the tip of an
Atomic Force Microscope~(AFM) to bend anchored MWNT's while simultaneously
recording the force exerted by the tube as a function of the displacement
from its equilibrium position, information from which the Young's modulus
of the nanotube can be extracted. Wong {\em et al.\/} have reported a mean
value of 1.28~TPa, which is in good agreement with the previous experimental
results. Also
Salvetat and coworkers~\cite{salvetat} have used a similar idea, which consists 
of depositing MWNT's on an ultra-filtration membrane. Many tubes are then
found to lie across the holes present in the membrane, with a fraction of 
their length suspended. The tip of an AFM is then used to exert a load 
on the suspended length of the nanotube, measuring at the same time the
nanotube deflection. The mean value of the Young's modulus obtained by
Salvetat~{\em et al.\/} was 0.81~TPa. A similar procedure has also been used
by Muster~{\em et al.\/}~\cite{muster}, who used an AFM to record the profile of
a MWNT lying across an electrode array. By assuming a simple Van der Waals
interaction law between the tube and the substrate, and regarding the 
nanotube as an elastic beam, the measured profile was found to be consistent
with a Young's modulus of approximately 1~TPa.

Other experiments, which as yet have not aimed at the mechanical 
characterization of nanotubes, nevertheless hint at other possible ways 
in which this characterization could be carried out. Among these we could
cite the embedding of nanotubes in resins~\cite{ajayan}, or measuring the
bending of anchored nanotubes in controlled magnetic fields~\cite{walter}.

All these experiments have contributed to confirming that nanotubes, both
SW and MW, have indeed exceptional mechanical properties, but there are
still questions that remain unresolved. The work of Chopra and 
Zettl~\cite{zettl} indicates that BN nanotubes are close in stiffness to 
carbon nanotubes, but the experimental error bars are too large to 
state categorically that one type of tube is stiffer than the other. The 
results of Salvetat {\em et al.\/}~\cite{salvetat} clearly indicate that
MWNT's synthesized by the arc-discharge method~\cite{arc}
are much stiffer than those produced by the catalytic decomposition of 
hydrocarbons~\cite{silvie}, which have a Young's modulus in the range of
10-50~TPa. This large difference is presumably a reflection of the influence
of the high density of defects present in the structure of the latter tubes, 
but a detailed quantification of the influence of defects on the mechanical 
properties of nanotubes is as yet missing.

\section{Previous Theoretical Work}
\label{sec:theoretical}

The mechanical properties of nanotubes have been addressed also by means
of theoretical calculations in a number of 
publications~\cite{iijima:bernholc,robertson,ruoff,molina,yakobson,nardelli,cornwell,lu}.
Most of these studies have been carried out using empirical potentials,
although tight-binding based models have also been occasionally used~\cite{molina}.
Though well-tested empirical potential models exist for carbon-based
systems~\cite{tersoff,brenner,matila}, to our knowledge no such model
exists for the composite systems, and therefore these materials have
been mostly studied using first-principles 
methodologies~\cite{rubio:corkill:cohen,miyamoto}. However, these latter
studies have
concentrated largely on electronic, structural and vibrational properties
of the composite systems, without addressing the issue of their
mechanical properties.

Lu~\cite{lu} has reported an extensive study of the Young's modulus, 
Poisson ratio and elastic constants of carbon nanotubes (both SW and MW, 
as well as ropes of SWNT's) using an empirical pair potential.
Yakobson {\em et al.\/}~\cite{yakobson} and Nardelli 
{\em et al.\/}~\cite{nardelli} have studied the behaviour of 
nanotubes subject to large axial strains.

\section{Model and Calculations}
\label{sec:model_and_calcs}

For the majority of the calculations reported here we have used a non-orthogonal
Tight-Binding scheme due to Porezag and coworkers~\cite{porezag}. Tight-Binding
(TB) methods~\cite{tb:review} lie in the centre region of the spectrum of 
simulation methods in both computational cost and reliability. Empirical
potentials~\cite{tersoff} are much cheaper to use, but their accuracy and
reliability is often questionable. First-principles methods~\cite{pw:review}
on the other hand are more reliable, but their computational demands are
orders of magnitude larger than for TB calculations, and often this makes their
use impractical. 

The TB model of Porezag {\em et al.\/}~\cite{porezag} contains two contributions
to the total energy: a so-called {\em band-structure\/} energy term, and a 
repulsive pair-potential. The band-structure energy is calculated as the sum
of the eigenvalues of the occupied states of a TB Hamiltonian. A matrix
representation of this Hamiltonian is constructed using a minimal basis set 
consisting of a single atomic-like orbital per atomic valence state. The
matrix elements are evaluated in the framework of Density Functional
Theory~(DFT) (normally in the Local Density Approximation, LDA), but 
retaining only two-centre contributions to the integrals. This means that
each matrix element of the Hamiltonian depends only on the relative distance
of the two atoms on which the corresponding basis functions are centred and 
the direction cosines of the internuclear vector~\cite{slater:koster,tb:review}.
Since the basis set is not orthogonal, the calculation of the band structure
energy requires the solution of a generalized eigenvalue 
problem~\cite{numerical_recipes}. The repulsive pair potential is then 
constructed in such a way that the total TB energy of a reference system
(usually the dimer) matches that of the full DFT calculation with the same
basis set. For more details on the construction of the model the reader should
consult the original references~\cite{porezag}, but an important point 
worth emphasizing here is the fact that no information concerning the mechanical
properties of the system under study are used in the construction of the
TB parametrisation.

Even thought the TB model used here is not fitted to any empirical data, it 
is nevertheless approximate and less accurate than conventional 
first-principles methods. For this reason we have also carried out
plane-wave (PW) pseudopotential DFT calculations for the (6,6) C and BN nanotubes,
in order to have the possibility of comparing our TB results with 
fully {\em ab initio\/} calculations. The PW calculations were performed
using Troullier-Martins pseudopotentials~\cite{troullier:martins} with a 
PW cutoff of 40~Ry for the basis set, and 10 reciprocal space points 
generated according to the Monkhorst-Pack scheme~\cite{monkhorst:pack}
to sample the one-dimensional Brillouin zone. The TB calculations used
$\Gamma$-point sampling only, but the periodic cells were chosen large enough
as to ensure the same degree of convergence in total energy differences
as were achieved in the PW calculations.

As we have seen in Sec.~\ref{sec:experimental}, the central 
property characterizing the stiffness
of nanotubes, to which the experiments have access, is the Young's modulus.
In bulk 3-D systems $Y$ is given by the following expression:
\begin{eqnarray}
Y = \frac{1}{V_0}
\left(\frac{\partial^2 E}{\partial \epsilon^2} \right)_{\epsilon=0},
\label{eq:young}
\end{eqnarray}
where $E$ is the total energy, $\epsilon$ is the strain and $V_o$ is the
equilibrium volume. The second derivative measures how rapidly the energy 
grows as the system is distorted out of its equilibrium configuration.
Usually $Y$ is given in units of pressure, and this is why the factor
of $V_0^{-1}$ appears in this formula. However, Eq.~({\ref{eq:young}) 
presents an ambiguity in the case of SWNT's, which stems
from the definition of $V_0$. To define $V_0$ for a SWNT one needs to
specify the wall thickness, and there is no clear way to define the 
thickness of a wall one-atom thick. In previous publications different 
authors have used different values for the tube
wall thickness, thought the most common convention has been to adopt the
value of the interlayer spacing in graphite. Nevertheless the value of $Y$
depends on the inverse of the wall thickness $\delta R$, and is therefore 
rather sensitive to the chosen value. In a recent 
publication~\cite{prl} we proposed a way to bypass this problem by using
an alternative definition of the Young's modulus, more appropriate to the
case of SWNT's:
\begin{eqnarray}
Y_s = \frac{1}{S_0} \left( \frac{\partial^2 E}{\partial \epsilon^2} 
\right)_{\epsilon=0}.
\label{eq:our_young}
\end{eqnarray}
Here $S_0$ is the surface area defined by the nanotube at zero strain, which
is a well defined quantity. Given that $V_0 = S_0 \delta R$, one can
recover the usual definition by simply dividing by $\delta R$:
$Y = Y_s / \delta R$, if one wishes to adopt a particular convention.

Another mechanical property of interest is the {\em Poisson ratio\/}, 
$\sigma$, which is defined by
\begin{eqnarray}
\frac{R - R_{eq}}{R_{eq}} = -\sigma \epsilon,
\label{eq:poisson}
\end{eqnarray}
where $R$ is the radius of the tube at strain $\epsilon$, and $R_{eq}$ is the
equilibrium (zero strain) tube radius. The Poisson ration measures how much
the tube contracts (expands) radially when subject to a positive (negative)
axial strain $\epsilon$.

We have performed a series of calculations using the Tight Binding model
discussed above aimed at determining $Y_s$ and $\sigma$ of both
carbon and composite SWNT's,including BN, $\mbox{BC}_3$ $\mbox{BC}_2\mbox{N}$
and $\mbox{C}_3\mbox{N}_4$ nanotubes. Two different graphite-like
$\mbox{BC}_2\mbox{N}$ structures are possible, but in our studies we have 
only considered the structure known as II, since this is the one
reported to be most stable~\cite{renata}.  

The calculations consist of taking
a section of a nanotube using periodic boundary conditions to simulate
an infinite tube, and subject it to both negative and positive strains
along the axial direction. At each strain the positions of all atoms in the
repeat-cell are fully relaxed without constraints, using the Conjugate
Gradients minimisation technique~\cite{numerical_recipes}. A given calculation
was assumed to have converged once the total energy varied less than 
$10^{-5}$~Hartree between two successive iterations. From these calculations
we obtained the total energy and atomic positions as a function of the 
axial strain imposed on the nanotube, from which we could calculate Young's 
modulus and the Poisson ratio, as well as the equilibrium structures of 
each nanotube considered.

\section{Results and Discussion}
\label{sec:results}

Let us first consider the difference of energy between a nanotube structure and 
the corresponding infinite flat graphene sheet. This energy difference
is known as the {\em strain energy\/} $E_s$, and we have plotted it in
Fig.~\ref{fig:strain_energy} for C, BN and $\mbox{BC}_3$ (n,n) nanotubes,
as a function of the tube diameter. It can be seen from 
Fig.~\ref{fig:strain_energy} that the strain energy varies as
$D^{-2}$, where $D$ is the tube diameter; we have performed
fits of functions of the form $a D^{-b}$ to the data, and the values
for the parameters $a$ and $b$ are given in Table~\ref{table:strain_energy}.

\begin{figure}
\begin{center}
\leavevmode
\epsfxsize=8cm
\epsffile{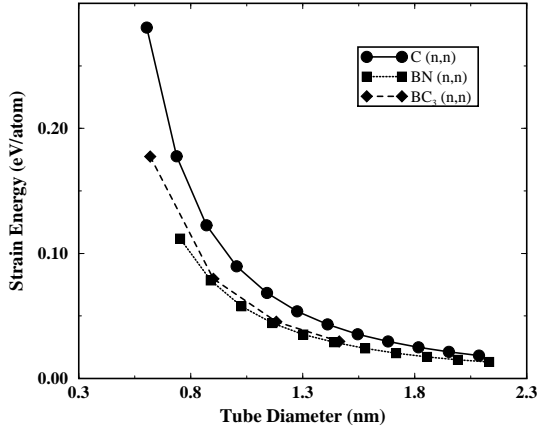}
\end{center}
\caption{Curvature strain energy as a function of the equilibrium tube
diameter, as obtained from the tight-binding calculations, for C, BN
and $\mbox{BC}_3$ nanotubes.}
\label{fig:strain_energy}
\end{figure}

That the strain energy should decay as $D^{-2}$ was predicted by
Tibbetts~\cite{tibbetts} (see also Mintmire and 
White~\cite{mintmire:white_in_ebbesen}) on the basis of continuum elasticity
theory, according to which the strain energy per atom is given by
\begin{eqnarray}
\frac{E_s}{N} = \frac{Y a^3}{6} \frac{\Omega}{D^2},
\label{eq:strain_energy}
\end{eqnarray}
where $Y$ is the Young's modulus of the tube, $a$ is a constant of the order
of the inter-layer spacing in graphite and $\Omega$ is the area per atom.
Note that from Eq.~(\ref{eq:strain_energy}) and the numerical fits to the strain
energy data of Fig.~\ref{fig:strain_energy} given in 
Table~\ref{table:strain_energy}, one can predict that the Young's modulus
of SWNT's of BN and $\mbox{BC}_3$ of a given diameter should be approximately
0.68 and 0.71 respectively that of a carbon nanotube of the same diameter.
As we shall see below, direct calculations of Young's modulus for these
tubes obey approximately this relation.

\begin{table}[t]
\begin{center}
\begin{tabular}{cccc}
$\mbox{B}_x\mbox{C}_y\mbox{N}_z$ & (n,m) & $a \times 10^2$ 
(eV $\mbox{nm}^2$/atom) & $b$ \\
\hline
C & (n,n) & 8.1 & 2.083 \\
  & (n,0) & 8.7 & 1.996 \\
BN & (n,n) & 5.5 & 1.984 \\
   & (n,0) & 5.6 & 1.980 \\
$\mbox{BC}_3$ & (n,n) & 5.8 & 1.984 \\
              & (n,0) & 5.6 & 2.048 \\
\end{tabular}
\end{center}
\caption{Parameters obtained from fitting the strain energy curves of
Fig.~\ref{fig:strain_energy} to a function of the form $a D^{-b}$. Note
that the value of $b$ is very close to 2 in all cases.}
\label{table:strain_energy}
\end{table}

\begin{table}[t]
\begin{tabular}{cccccc}
$\mbox{B}_x\mbox{C}_y\mbox{N}_z$ & (n,m) & $D_{eq}$~(nm) & $\sigma$ & 
$Y_s$~($\mbox{TPa}\cdot\mbox{nm}$) & $Y$~(TPa) \\
\hline
C & (10,0) & 0.791 & 0.275 & 0.416 & 1.22 \\
  & (6,6) & 0.820 & 0.247 & 0.415 & 1.22 \\
  &       & (0.817)      &       & (0.371) & (1.09) \\
  & (10,5) & 1.034 & 0.265 & 0.426 & 1.25 \\
  & (10,7) & 1.165 & 0.266 & 0.422 & 1.24 \\ 
  & (10,10) & 1.360 & 0.256 & 0.423 & 1.24 \\
  & (20,0) & 1.571 & 0.270 & 0.430 & 1.26 \\
  & (15,15) & 2.034 & 0.256 & 0.425 & 1.25 \\ \hline
BN & (10,0) & 0.811 & 0.232 & 0.284 & 0.837 \\
   & (6,6) & 0.838 & 0.268 & 0.296 & 0.870 \\
   &       & (0.823)      &       & (0.267) & (0.784) \\
   & (15,0) & 1.206 & 0.246 & 0.298 & 0.876 \\
   & (10,10) & 1.390 & 0.263 & 0.306 & 0.901 \\
   & (20,0) & 1.604 & 0.254 & 0.301 & 0.884 \\ 
   & (15,15) & 2.081 & 0.263 & 0.310 & 0.912 \\ \hline
$\mbox{BC}_3$ & (5,0) & 0.818 & 0.301 & 0.308 & 0.906 \\
              & (3,3) & 0.850 & 0.289 & 0.311 & 0.914 \\
              & (10,0) & 1.630 & 0.282 & 0.313 & 0.922 \\ 
              & (6,6) & 1.694 & 0.279 & 0.315 & 0.925 \\ \hline
$\mbox{BC}_2\mbox{N}$ II & (7,0) & 1.111 & 0.289 & 0.336 & 0.988 \\
                         & (5,5) & 1.370 & 0.287 & 0.343 & 1.008 \\ \hline
$\mbox{C}_3\mbox{N}_4$   & (6,0) & 0.913 & 0.280 & 0.192 & 0.565 \\
                         & (8,0) & 1.210 & 0.238 & 0.207 & 0.610 \\
                         & (6,6) & 1.558 & 0.177 & 0.228 & 0.670 \\
                         & (8,8) & 2.075 & 0.132 & 0.233 & 0.684 \\
\end{tabular}
\caption{Structural and elastic properties of selected nanotubes obtained
from the tight-binding calculations reported here. Young's modulus 
values given in parenthesis were obtained from first-principles calculations.
Also the value of $Y$ 
with the convention $\delta R = 0.34$~nm is given for comparison.} 
\label{table:tb_properties}
\end{table}

A first indication that the TB model used in this work is a reliable one
stems from the good agreement obtained in the strain energy as calculated
here and that calculated from first-principles methods and reported 
elsewhere~\cite{rubio:corkill:cohen,miyamoto}. Another indication comes
from the fact that a certain buckling on the surface of the BN nanotubes is 
predicted to occur, also in agreement with preliminary
first-principles calculations~\cite{rubio:corkill:cohen}. This buckling, which
results from the B atoms displacing inwards towards the tube axis,
while the N atoms displace in the opposite direction, is a consequence of the
slightly different hybridizations of B and N on the curved surface of 
the nanotube. The amount of buckling is dependent on the tube diameter, 
but it is otherwise independent of the tube structure for arm-chair and 
zig-zag nanotubes.

In Table~\ref{table:tb_properties} 
we give the obtained values of structural and 
mechanical properties for a set of nanotubes obtained from our calculations.
For comparison we also give results for the (6,6) C and BN nanotubes calculated
with PW pseudopotential DFT calculations. As can be seen, the agreement between
the TB and first-principles calculations in both structural and mechanical
properties is rather good. Notice also that the values of $Y_s$ (and in fact
those of $Y$ also) for C, BN and $\mbox{BC}_3$ nanotubes of similar
diameters are approximately in the same ratio as predicted from 
Eq.~(\ref{eq:strain_energy}).

\begin{figure}[t]
\begin{center}
\leavevmode
\epsfxsize=8cm
\epsffile{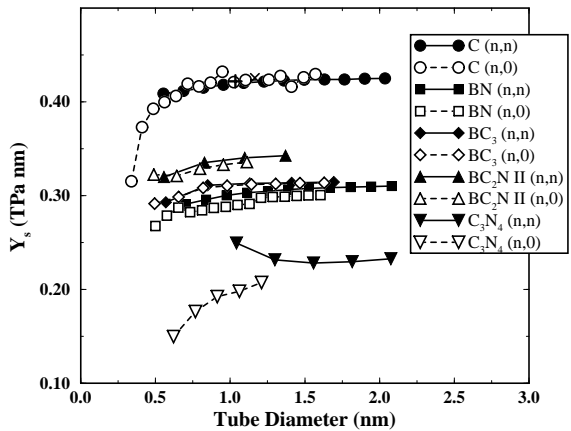}
\end{center}
\caption{Young's modulus as a function of the tube diameter for C, BN,
$\mbox{BC}_3$, $\mbox{BC}_2\mbox{N}$ (structure II only) and 
$\mbox{C}_3\mbox{N}_4$, as
calculated from the tight-binding simulations.
Results obtained for (n,n) nanotubes (filled symbols), (n,0) nanotubes
(empty symbols) and also for C (10,5) ($+$) and (10,7) ($\times$) are shown.}
\label{fig:young}
\end{figure}

\begin{figure}[t]
\begin{center}
\leavevmode
\epsfxsize=8cm
\epsffile{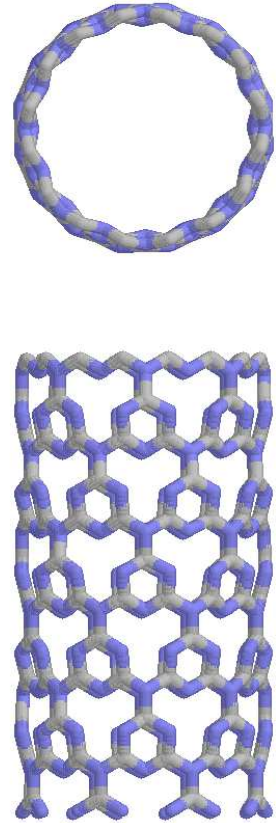}
\end{center}
\caption{Relaxed structure for the $\mbox{C}_3\mbox{N}_4$ (8,0) 
nanotube, as obtained from the TB calculations. The dark shaded atoms are
Nitrogen atoms, while the lighter ones are Carbon atoms.}
\label{fig:c3n480}
\end{figure}

In Fig.~\ref{fig:young} the values $Y_s$ have been plotted as a function
of the tube diameter for the different types of tubes considered in this 
work. The first observation that can be extracted from 
Table~\ref{table:tb_properties} and Fig.~\ref{fig:young} is the fact that
carbon nanotubes are predicted to have the highest Young's modulus of
all the different types of tubes considered. BN and $\mbox{BC}_3$ tubes
have very similar values of $Y_s$, though the latter have slightly larger
values. The $\mbox{BC}_2\mbox{N}$ nanotubes are predicted to be slightly
stiffer than the BN and $\mbox{BC}_3$ tubes, while $\mbox{C}_3\mbox{N}_4$
nanotubes lie well below the rest in stiffness. 
The value of $Y_s$ we
obtain for the wider C nanotubes, 0.43~TPa~nm, corresponds to a Young's modulus
of 1.26~TPa in the conventional definition of Eq.~(\ref{eq:young}), if
we take $\delta R = 0.34$~nm, i.e. the inter-layer spacing in graphite.
This value is in very good agreement with the experimental value
recently obtained by Krishnan and coworkers~\cite{krishnan} for SWNT's
(1.25~TPa). It is also in rather good agreement with the value of
1.28~TPa reported by Wong {\em et al.\/}~\cite{lieber}, though this later
value was obtained for MWNT's. However, it is expected that the Young's 
modulus be mostly determined by the intra-wall C-C bonds, and it is therefore
not surprising that the values look so similar for both MW and SWNT's.
For composite nanotubes, the only experimental data on mechanical 
properties currently available to our knowledge are the results of 
Chopra and Zettl~\cite{zettl}, who have measured $Y$ for BN MWNT's.
They quote a value of 1.22~TPa, which is somewhat larger than the result
we obtain for these tubes, but nevertheless the agreement is close.
Tough it may seem that the choice $\delta R = 0.34$~nm is somewhat arbitrary,
it should be pointed out that the experimental results are not free of
this arbitrariness either, given that to obtain a value of the Young's
modulus according to Eq.~(\ref{eq:young}), it is necessary to interpret 
the experimental observations on the basis of some mechanical model, usually
a hollow cylinder with a certain wall thickness. Clearly, in the case of
SWNT's the question of how to choose $\delta R = 0.34$~nm applies 
to experiments as well as to theoretical calculations. 

\begin{figure}[t]
\begin{center}
\leavevmode
\epsfxsize=8cm
\epsffile{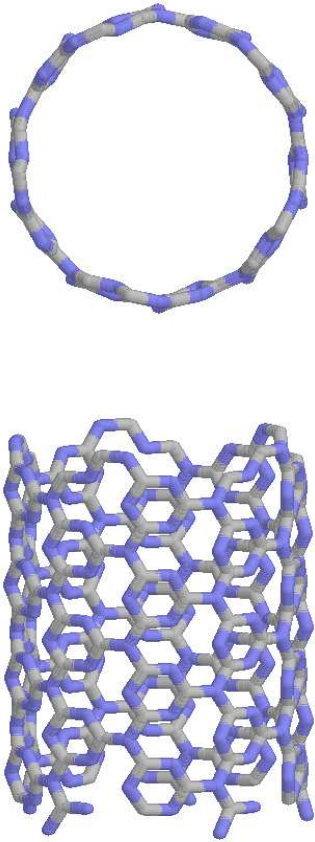}
\end{center}
\caption{Relaxed structure for the $\mbox{C}_3\mbox{N}_4$ (5,5) 
nanotube.}
\label{fig:c3n455}
\end{figure}

The reason why $\mbox{C}_3\mbox{N}_4$ nanotubes are predicted to be 
so much softer than 
all other types of nanotubes considered in this work is the fact that for
a given amount of tube surface, tubes of this composition have a smaller
density of chemical bonds. Indeed, these tubes present a more hollow
structure when compared to the other, perfectly hexagonal nanotubes 
(see Figs.~\ref{fig:c3n480} and~\ref{fig:c3n455}). 
It is also interesting to note that,
while C, BN, $\mbox{BC}_3$ and $\mbox{BC}_2\mbox{N}$ nanotubes do not show
noticeable differences for the structural or mechanical properties as
the chiral angle is varied, sizeable differences are observed in the
case of $\mbox{C}_3\mbox{N}_4$ nanotubes. It can be seen in
Table~\ref{table:tb_properties} that the (n,n) $\mbox{C}_3\mbox{N}_4$
tubes have a higher Young's modulus than the (n,0), though both seem
to converge towards the same number in the limit of large diameters.
This difference is also reflected in the structure, as can be seen 
in Figs.~\ref{fig:c3n480} and~\ref{fig:c3n455}. Notice 
how the the hexagons are at an angle
with the surface of the tube which is different in (n,n) and (n,0) 
tubes. 

As for comparison with other theoretical predictions, the results quoted
by Lu~\cite{lu} are somewhat smaller than ours. The results in 
ref.~\cite{lu} are 0.97~TPa for all tubes. This
difference is most likely due to the different approaches (empirical
potentials and TB) used in that work and ours. We also observe in our
results for $Y_s$ a slight dependence on the tube diameter. As the 
diameter becomes larger, $Y_s$ approaches a plateau value which corresponds
to the value calculated for the flat graphene-like sheet of each nanotube
composition. Interestingly, the approach to the limit value is from 
below, as can be expected if one considers that bending a flat graphene
sheet weakens the bonds. Given that it is the strength of the chemical bonds
which determines the actual value of the Young's modulus, it is natural that 
small-diameter (high curvature) tubes have smaller Young's moduli, and in
the limit of large diameters, the mechanical properties essentially correspond
to those of the flat graphene sheet. In contrast, the results of Lu~\cite{lu}
are largely insensitive to the tube diameter. This is due to the
fact that pair-potential models, such as the one used by Lu, 
do not reflect the changing nature of the chemical bonding as the 
curvature is changed. To reproduce this effect, a model sensitive to the
changing environment (i.e. a {\em many-body\/} model) is required.

To summarize, we have used a non-orthogonal TB model parametrised for 
C, B and N based systems to perform a systematic study of the energetic,
structural and mechanical properties of single-wall nanotubes of
different chemical composition. We have
checked the accuracy of our predictions against some first-principles
calculations, and the agreement obtained is good. Furthermore, we obtain
good agreement with the available experimental data. We obtain 
strain energy vs. diameter curves which obey very closely the 
expected $D^{-2}$ behaviour. Our results 
show that carbon nanotubes are expected to be stiffer than any of the composite
nanotubes considered, having a Young's modulus of approximately 1.3~TPa,
which corresponds to that of a flat graphene sheet within the same 
theoretical model. The $\mbox{C}_3\mbox{N}_4$ nanotubes, which present 
a more hollow structure than the other tubes, are predicted to have
a Young's modulus nearly half that of the carbon nanotubes.

We are grateful to G.~Seifert and T.~Heine for providing the TB 
parametrisations used in this work. J.A.~Alonso, M.J.~L\'{o}pez and 
M.~Galtier are gratefully acknowledged for helpful discussions. 
This work was carried out within the framework of the 
EU Transfer and Mobility of Researchers Namitech project under contract
No. ERBFMRX-CT96-0067 (DG12-MITH) and Grant No. DGIS-PB95-0202
of the Spanish Ministry of Education. The use of computer facilities
at C4 (Centre de Computaci\'{o} i Comunicacions de Catalunya) 
and CNUSC (Montpellier) is also acknowledged.

\end{document}